\documentclass[letterpaper, 10pt, conference,twocolumns]{IEEEtran}
\makeatletter
\def\ps@headings{%
\def\@oddhead{\mbox{}\scriptsize\rightmark \hfil \thepage}%
\def\@evenhead{\scriptsize\thepage \hfil \leftmark\mbox{}}%
\def\@oddfoot{}%
\def\@evenfoot{}}
\makeatother
\usepackage{graphics} % for pdf, bitmapped graphics files
\usepackage{epsfig} % for postscript graphics files
\usepackage{times} % assumes new font selection scheme installed
\usepackage{amsmath} % assumes amsmath package installed
\usepackage{amssymb}  % assumes amsmath package installed
\usepackage{amsthm}\usepackage{amsthm}
\usepackage[english]{babel}
\usepackage{verbatim}
\usepackage{subfigure}
\usepackage{url}
\usepackage{float}
\usepackage[small]{caption}
\usepackage{graphicx}
\usepackage{color}
\usepackage{flushend}
\newtheorem{defi}{Definition}
\newtheorem{prob}{Problem}
\newtheorem{lemma}{Lemma}
\newtheorem{cor}{Corollary}

\newtheorem{thm}{Theorem}
\newtheorem{remark}{Remark}
\newcommand{\umax}{u_{\max}}
\newcommand{\umin}{u_{\min}}
\newcommand{\U}{\mathcal U}
\newcommand{\R}{\mathbb R}
\newcommand{\dVx}{DV}
\newcommand{\no}{\nonumber}

\newcommand{\bu}{\mathbf u}

\DeclareMathOperator*{\argmax}{\displaystyle \arg\!\max}

\newcommand{\fdp}[1]{\textcolor{black}{{#1}}}
 \newcommand{\QED}[1]{\IEEEQEDoff \hspace{#1mm} \IEEEQEDhere}
 \newcommand{\noi}{\noindent} 
\IEEEoverridecommandlockouts
\renewcommand{\baselinestretch}{.98}

\begin{document}

\title{Differential Games of Competition in Online Content Diffusion}
\author{Francesco De Pellegrini$^\diamond$\thanks{$^\diamond$ CREATE-NET, via Alla Cascata 56 c, 38100 Trento, Italy; INRIA Sophia-Antipolis, 2004 Route des Lucioles, 06902 Sophia-Antipolis Cedex, France}, Alexandre Reiffers$^\star$  and Eitan Altman$^\star$}% and Tamer Ba\c sar}

\maketitle

\begin{abstract}
Access to online contents represents a large share of the Internet traffic. Most 
such contents are multimedia items which are user-generated, i.e., posted 
online by the contents' owners. In this paper we focus on how those who provide 
contents can leverage online platforms in order to profit from their large
base of potential viewers. 

Actually, platforms like Vimeo or YouTube provide tools to 
accelerate the dissemination of contents, i.e., recommendation lists and other re-ranking mechanisms. 
Hence, the popularity of a content can be increased by paying  a cost for advertisement: doing so, it 
will appear with some priority in the recommendation lists and will be accessed more frequently by the platform users.  

Ultimately, such acceleration mechanism engenders a competition 
among online contents to gain popularity. \fdp{In this context, our focus
is on the structure of the acceleration strategies which a content 
provider should use in order to optimally promote a content given a 
certain daily budget. Such a best response indeed depends on the strategies 
adopted by competing content providers. Also, it is a function of the potential 
popularity of a content and the fee paid for the platform advertisement service.}

We formulate the problem as a differential game and we solve it
for the infinite horizon case by deriving the structure of certain Nash 
equilibria of the game.
\end{abstract}

\maketitle
%%%%%%%%%%%%%%%%%%%%%%%%%%%%%%%%%%%%%%%%%%%%%%%%%%%%%%%%%%%%%%%%%%%%%%%%%%%%%%%%%%%%%%%%%%%%%%%%%%%%%%%%%%%%%%%%%%%%%%%%
%%%%%%%%%%%%%%%%%%%%%%%%%%%%%%%%%%%%%%%%%%%%%%%%%%%%%%%%%%%%%%%%%%%%%%

\section{Introduction}\label{sec:1}

%%%%%%%%%%%%%%%%%%%%%%%%%%%%%%%%%%%%%%%%%%%%%%%%%%%%%%%%%%%%%%%%%%%%%%
%%%%%%%%%%%%%%%%%%%%%%%%%%%%%%%%%%%%%%%%%%%%%%%%%%%%%%%%%%%%%%%%%%%%%%%%%%%%%%%%%%%%%%%%%%%%%%%%%%%%%%%%%%%%%%%%%%%%%%%%

Online content delivery represents an ever increasing fraction of Internet traffic. In the case of online videos, 
the support for content distribution is provided by commercial platforms such as Vimeo or YouTube. In many 
cases, such contents are also delivered by means of social network platforms. One core feature of such systems
is the delivery of user-generated content (UGC): platform users become often producers of the contents which 
populate those systems. 

Reference figures for UGC platforms are indeed those of YouTube, with over $6$ billion hours of videos watched each month, 
 which averages as an hour for every person on Earth a month. New UGCs are continuously created: $100$ hours of video 
are uploaded every minute by the YouTube platform's users\footnote{{\tt http://www.youtube.com/t/press\_statistics/}}. 

A relevant parameter for the UGC platforms owners is the {\em viewcount}, i.e., the number of times an item has 
been accessed. Viewcount in fact represents one of the possible metrics to measure content popularity. In turn, a 
popular video becomes a source of revenue because of  click-through rates of linked advertisements: those 
are actually part of  the YouTube's business model. 

Among several research works in the field, many efforts have been spent to characterize the dynamics 
of popularity of online media contents~\cite{ChaUtube,crane2008viral,Gill07youtubetraffic,RatkiewiczBurstyPoP,ChatFirstStep,ChaTON}. 
The ultimate target there would be indeed to provide models able to perform the early-stage prediction of a content's popularity~\cite{SzaboPop}. 

Such studies have highlighted certain phenomena that are typical of UGC
delivery. %This includes the fact that a significant share of content
%gets basically no views~\cite{ChaTON}, as well as the fact that popularity may see some
%bursts, when content ``goes viral''~\cite{RatkiewiczBurstyPoP}. 
A key study in~\cite{SzaboPop}  shows that the dynamics of popularity of online contents
experiences two phases. In the initial phase, a content gains popularity through
advertisement and other marketing tools. Afterwards, UGC platform mechanisms  
induce users to access contents by re-ranking mechanisms. Those also appear to be 
main drivers of popularity. 

Motivated by such findings, we model the behavior of those who create content --
shortly content providers in the rest of the paper -- as a dynamic game. Once they 
generate a content, in particular, they leverage on UGC platforms to diffuse it. We 
note that, by paying a fee for the advertisement service of the UGC platform, a content 
provider is able to receive a preferential treatment to her content such in a way that the 
rate of propagation is increased. Clearly, this engenders a competition among content providers to 
capture the attention of potential viewers at faster rate than other contents. 

\begin{figure}[t]{
  \begin{center}
      \includegraphics[width=0.5\textwidth]{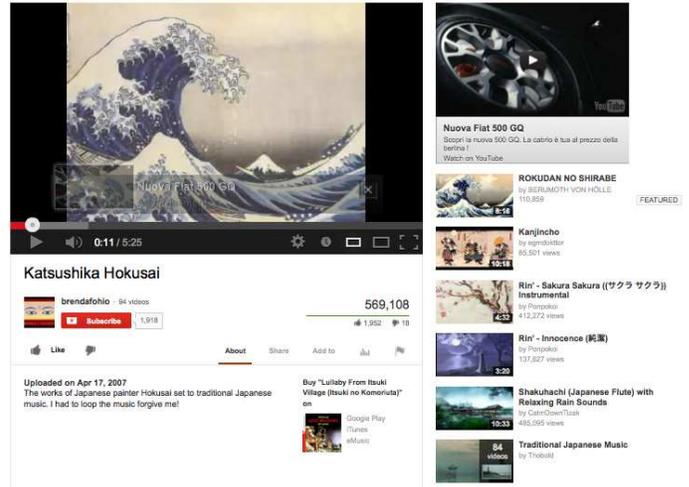}
%      \put(-210,120){$X(t,\theta)$}
%      \put(-110,10){$t_\beta(\theta)$}\put(-30,10){$\tau$}\put(-80,5){$\tau-t_\beta(\theta)$}  
%      \put(-120,110){$\beta$}
\end{center}     
  \caption{A sample video: the icons on the right of the main window is the recommendation list. The upper entry 
  seen there is a commercial advertisement, whereas the second entry is tagged "Featured". The remaining entries 
  are sorted according to their viewcount.}\label{fig:hakusay}}
\end{figure}

To this respect, the notion of {\em acceleration} is a key concept. An example reported 
in Fig.~\ref{fig:hakusay} explains how a video can be accelerated by  the UGC platform.
In our example we performed a generic search {\em ``Hakusai''}  which produces a series of 
output results for matching contents. The one reported in the figure is one with viewcount 
$568507$ that has been listed by the main YouTube. 

In particular,  Fig.~\ref{fig:hakusay}  represents the viewer's screen: in the central part of 
the window it stands  the video. However, there exists a recommendation list on the right
 as provided by  the platforms' search engine.  

It is important to observe the two videos recommended on the top of the list. The first one is an advertisement 
of a known commercial activity. In order to appear in the top position of the list,  that content 
has been paying a fee to the UGC platform owner. In the second position, a link appears to a video 
which is tagged {\em featured}. The meaning  of the term featured is that the video linked there 
was placed high on the recommendation list either because it is a very popular video or because
it is a  partner video. A partner video, as in the case of advertisements, is from someone 
who pays a fee to rank higher in the recommendation list. The other videos in the 
recommendation list are ranked according to the default order, e.g., the viewcount. 
Another advertisement with the suggestion to buy a product is appearing at the bottom of the figure. 
 \fdp{In this paper we focus on the acceleration of featured videos. In fact, in order to accelerate 
 a video, customers perform a {\em promoted video campaign} on YouTube; to do so, content 
 providers are required four steps: choose a video, attach promotional text, some keywords 
 by which the promotion is performed, and set the  {\em daily budget} amount allowed. }

\fdp{With respect to the acceleration cost, it is important  to note that the so called pay-per-view model 
is applied. I.e., the YouTube pay-per-view policy for acceleration is meant to charge the content provider a 
fixed amount each time a viewer has accessed the content. Charging is triggered by a click-through 
on the icons of the promoted content which appear in the recommendation list. However, for the platform 
owner it is best that the customer's daily budget is attained. Then, the total cost paid in order to increment 
the number of views would increase linearly in time. A linear cumulative cost for acceleration is also one of our 
assumptions in the rest of the paper.}

Now, since the viewer' browser has finite size, only those who are able to appear in the higher end 
of the recommendation list are visible without scrolling. Thus,  those are  accessed with higher
probability: the viewcount of a content is expected  indeed to grow faster, i.e., to be accelerated, 
 whenever it is showed higher in the list. 

In this work, we consider a competition between several contents. The promotion fees, i.e., the 
cost to accelerate the viewcount, will depend on the content provider and may depend on the content itself. Even 
the rate of propagation, i.e., the rate at which viewers access the content, may depend on the content.  Finally, 
%based on the popularity level of a content, on the potential size of the interested audience as well as the
%number of past visualizations, i.e., the content's viewcount, 
each content provider may decide whether or not
to purchase priority to accelerate the popularity of the content for a certain period. 

The objective of this paper is to determine the best strategy for a content provider in order to accelerate a content
 and study the resulting equilibria of the system.  To this aim, we propose a game theoretical framework 
 rooted in differential games. The solution of the problem allows us to provide guidelines for the advertisement 
 strategies of content providers.

A brief outline of the paper follows. In Sec.~\ref{sec:2} we revise the main results in literature for online content diffusion. 
In Sec.~\ref{sec:3} we introduce the system model and the differential game subject of this paper. In Sec.~\ref{sec:4} 
we derive the analysis of best responses whereas symmetric Nash equilibria are characterized in Sec.~\ref{sec:symplay}. 
In Sec.~\ref{sec:5} we tackle the limit case of small discounts and in the following Sec.~\ref{sec:6} we briefly touch the
analysis of the game  for a finite horizon. A section with conclusions and future directions ends the paper.

%%%%%%%%%%%%%%%%%%%%%%%%%%%%%%%%%%%%%%%%%%%%%%%%%%%%%%%%%%%%%%%%%%%%%%%%%%%%%%%%%%%%%%%%%%%%%%%%%%%%%%%%%%%%%%%%%%%%%%%%

%%%%%%%%%%%%%%%%%%%%%%%%%%%%%%%%%%%%%%%%%%%%%%%%%%%%%%%%%%%%%%%%%%%%%%%%%%%%%%%%%%%%%%%%%%%%%%%%%%%%%%%%%%%%%%%%%%%%%%%%
%%%%%%%%%%%%%%%%%%%%%%%%%%%%%%%%%%%%%%%%%%%%%%%%%%%%%%%%%%%%%%%%%%%%%%

\section{Related works}\label{sec:2}

%%%%%%%%%%%%%%%%%%%%%%%%%%%%%%%%%%%%%%%%%%%%%%%%%%%%%%%%%%%%%%%%%%%%%%
%%%%%%%%%%%%%%%%%%%%%%%%%%%%%%%%%%%%%%%%%%%%%%%%%%%%%%%%%%%%%%%%%%%%%%%%%%%%%%%%%%%%%%%%%%%%%%%%%%%%%%%%%%%%%%%%%%%%%%%%

The dynamics of popularity of online contents has been attracting attention from the 
research community. \cite{Gill07youtubetraffic} proposed an analysis of the 
YouTube system focusing on the characteristics of the traffic generated by that platform. 
\cite{ChatFirstStep}  addressed the relation between metrics used to evaluate popularity, 
such as number of comments, ratings, or favorites. In this paper our analysis is restricted 
to the viewcount. 

In \cite{RatkiewiczBurstyPoP} the authors  study the ranking change 
induced by UGC online platforms. %The model is meant to overcome the limitations of the preferential 
%attachment models: b
Bursty  acceleration in content's viewcount is found to depend on the way how 
online platforms expose popular contents to users and on re-ranking of existing contents. 
Here, we model the competition that arises when several content providers leverage 
such acceleration tools. 

In literature, competition in epidemic processes has been addressed with game theoretical tools. In \cite{Alon2010}, 
the authors focus on an economic game on graphs, where firms try to conquer the largest market share. They derive
the complexity for the computation of the equilibria of the game;  results for the price of anarchy 
in those games has been developed recently in \cite{Tzoumas2012}. 

Emergence of equilibria of the Wardrop type has been studied in \cite{NetSciCom} from viewer's perspective. 
Our objective here is to describe the content provider viewpoint in a dynamic game framework.
 
In \cite{Jiang13} the authors consider information propagation through social networks. The question there is 
how the finite budget of attention of individuals influences the rate at which contents can be pushed into 
the other players' network. In our work, we limit our focus to  the case of online content diffusion in UGC 
platforms. 

%The general context for the model developed in this paper is influenced by the work in~\cite{Altman2012}.

{\noindent \em Novel contributions:} In this paper, we provide a complete framework for the analysis of dynamic games in UGC provision. 
 Under a meanfield approximation for contents diffusion, differential games \cite{BasarDynGames} provide the model for capturing 
 the strategic behavior of competing content providers. Our main findings are:
\begin{itemize}
\item The structure of the best response of content providers and a method for calculating it; 
\item Conditions for the existence and uniqueness of symmetric Nash equilibria in threshold form;
\item Approximated asymmetric Nash equilibria in the regime of small discounts.
%\item The closed form expressions for the value function in the case of finite horizon.
\end{itemize}
To the best of the authors' knowledge, results on Nash equilibria for differential games in UGC provision have not been derived so far in literature. 

%%%%%%%%%%%%%%%%%%%%%%%%%%%%%%%%%%%%%%%%%%%%%%%%%%%%%%%%%%%%%%%%%%%%%%%%%%%%%%%%%%%%%%%%%%%%%%%%%%%%%%%%%%%%%%%%%%%%%%%%
%%%%%%%%%%%%%%%%%%%%%%%%%%%%%%%%%%%%%%%%%%%%%%%%%%%%%%%%%%%%%%%%%%%%%%

\section{System Model}\label{sec:3}

%%%%%%%%%%%%%%%%%%%%%%%%%%%%%%%%%%%%%%%%%%%%%%%%%%%%%%%%%%%%%%%%%%%%%%
%%%%%%%%%%%%%%%%%%%%%%%%%%%%%%%%%%%%%%%%%%%%%%%%%%%%%%%%%%%%%%%%%%%%%%%%%%%%%%%%%%%%%%%%%%%%%%%%%%%%%%%%%%%%%%%%%%%%%%%
\begin{table}[t]\caption{Main notation used throughout the paper}
\centering
\begin{tabular}{|p{0.10\columnwidth}|p{0.8\columnwidth}|}
\hline
{\it Symbol} & {\it Meaning}\\
\hline
$N$ & number of players (content providers) \\
%$M$ & number of potential viewers of a content \\
$\lambda_i$ & intensity of views per second for content $i$\\
$\tau$ & time horizon\\
$x_i(t)$ & fraction of viewers having viewed content $i$ at time $t$\\
$x(t)$ & summation $ \sum_i x_i(t)$\\
$y(t)$ & $y(t):=1-x(t)$ \\
$z_i$ & :=$x_i(0)$ will be taken 0 unless otherwise stated.\\
$u_i(t)$ & acceleration control (strategy) for player $i$; $\mathbf u=(u_1,u_2,\ldots,u_N)$ \\
             & $1\leq \umin < \umax < \infty$  \\
$\bu_{-i}(t)$ & strategy profile for all players not $i$ \\
$a$ & sum of the $\lambda_i u_i$ \\
$a_{-i}$ & sum of the $\lambda_j u_j$ for all players $j\not = i$\\
$p_i$ & discount factor for player $i$, $p_i>0$ \\
%$\xi_i$ & target fraction for small provider $i$\\
\hline
\end{tabular}\\[-5mm]
\label{tab:notation}
\end{table}

The main symbols used in the paper are reported in Tab.~\ref{tab:notation}.

In our system model we assume that $N$ competing content providers release a content each 
and viewers will access one of such contents at the earliest  chance.

We assume a base of $M$ potential viewers who can access each of the $N$ contents. To this respect, 
we adopt a fluid approximation which is assumed to hold for large $M$, and let content viewers to access 
content $i$ according to a point process with intensity $\lambda_i$.  I.e., content $i$ is accessed by a 
randomly picked viewer every $\lambda_i^{-1}$ seconds. 

In general $\lambda_i\not=\lambda_j$: in fact, contents may experience diverse popularity, and so different 
intensities. Every content provider will participate to the content diffusion in some time frame $[0,\tau]$. Also, 
$\tau=\infty$ in the development of the infinite horizon  formulation of the game.

\fdp{Also, we assume that the access is exclusive, i.e., viewers do not acquire another content 
after accessing a competing one. In general, the above assumption may appear restrictive. But, 
it does apply to several content types, e.g., the same episode of a series posted by different users, 
or a video related to a specific event such as a sport match. \\
More in general, our model applies to the case when the viewers of interest are 
those who access the content before other competing contents.}

Using the advertisement options of the platform, content providers can pay a cost in order to accelerate the diffusion of their 
video: viewers will access the content according to the intensity $u_i \lambda_i$, where $u_i \geq 1$
 is the acceleration control for player $i$. The maximum acceleration is bounded as $\umax$, and the minimum acceleration
 is $\umin$: $1\leq \umin < \umax$.

\fdp{Finally, there is a linear cost paid for the acceleration control: such a cost represents the ideal case, i.e., when a content 
provider receives per day a certain number of new views per cent paid to the platform owner. Conversely, 
the case with no acceleration, namely, $u_i=\umin$, falls back to the default intensity $\umin\lambda_i$. Of course, this happens  at zero cost.} 
%The acceleration control should be interpreted, as mentioned in the previous sections, 
% as a specific advertisement effort. In the case of the YouTube channels, this would be a cost which is paid to the channel owner to increase the ranking 
%in the recommendation list or to make links to the content visible in other target online resources.

We introduce below the game model  that we use to describe the competition 
among content providers in order to accelerate the dynamics of the viewcount. 
The formulation of the problem is initially provided in general to cover both in the finite 
horizon and in the infinite horizon case. Within the scope of the paper, most of the development 
is restricted to the infinite horizon analysis. 

%%%%%%%%%%%%%%%%%%%%%%%%%%%%%%%%%%%%%%%%%%%%%%%%%%%%%%%%%%%%%%%%%%%%%%

\subsection{Game model}\label{subsec:model}

%%%%%%%%%%%%%%%%%%%%%%%%%%%%%%%%%%%%%%%%%%%%%%%%%%%%%%%%%%%%%%%%%%%%%%
The differential game model for online content diffusion is composed as follows.\\
{\noi \em Players:} players are content providers, who compete in order to diffuse their content over a base of 
potential viewers. Since all players share the same base, the formulation will result in a competitive 
differential game.  \\
{\noi \em Strategies:} the strategy of each player is the acceleration control. The control is thus dynamic, since 
each player should determine at each point in time the acceleration $u_i(t)$. \\%Strategies are then drawn from the set $[0,\tau]^{[1,\umax]}$.\\
{\noi \em Utilities:} the utility for player $i$ is linear and has two terms. First, there is a cost paid for accelerating the content. 
Second, there is a revenue represented by the number of copies. The total utility is defined, as customary in differential 
games, as the integral of an instantaneous utility. 

We denote $x_i$ the fraction of  viewers who have accessed to the contents generated 
by the $i$-th content provider. The governing equation for the dynamics of the $i$-th content's viewcount is 
\begin{eqnarray}\label{e1}
 \dot x_i = \lambda_i u_i (1-x), \quad i=1,2,\ldots,N
\end{eqnarray} 
where $x= \sum_i x_i$ is the total fraction of viewers who accessed some content; the initial condition is $x_i(0)=z_i$. 
Actually, \eqref{e1} is a fluid approximation for the dynamics of the fraction of viewers of content $i$. %A detailed 
%discussion on the validity of such approximation for a similar controlled stochastic process is available in \cite{lucile} based on the work of 
%Bena\"im and Le Boudec in \cite{Benaim}.

\begin{remark}
\fdp{The fluid approximation which we use in this context can be justified formally with 
the derivation proposed by \cite{benaim}. In particular, let $ \widehat {\bf X}^{(M)}(t)$ be a $N$ 
dimensional vector whose components are $\widehat X_i^{(M)} (t)$, for $i=1,\dots,N$. Here, 
$\widehat X_i^{(M)} (t)$ stands for the fraction of the potential viewers that watched the content 
at time $t$, when the basin of users has size $M$: it represents the branching process of the 
$i$-th content being watched.  Thus, when we refer to fluid approximations that describe the 
dynamics  of the fraction viewers watching the content,  we are referring the meanfield 
approximation of such process. In particular, for a formal explanation of the convergence 
for large $M$ to the fluid approximations of the type used hereafter, the reader can  
refer to \cite{lucile}.}
\end{remark}

The acceleration control $u_i$, namely the strategy of player $i$, belongs to the space of the piecewise 
continuous functions $\U=\{u \in  \mbox{p.w.c. functions of \;}[\umin,\umax]^{[0,\tau]}\}$. 
 
Hence, because the control is upper bounded, the above ODE system 
\eqref{e1} is Lipschitz continuous, and because it 
is lower bounded, it is so uniformly in the control, so that a solution 
at large is guaranteed to exist unique for a given strategy profile $\bu=(u_1,\ldots,u_N)$ (\cite{BaCaDo}, pp. 99).

%The control function $u \in {\mathbb R}^{U}$, where $U=[\umin,\umax]$, 
%where $\umax>1$ is represents the positive upper bound for the 
%acceleration term. 
The cost function for the $i$-th player is given by
\begin{eqnarray}\label{e2}
J_i(x,u)=\int_0^{\infty} e^{-p_i s } \big(-\dot x_i(s) + \gamma_i \big ( u_i(s) - u_{min}\big) )ds\nonumber \\
           = \int_0^{\infty} e^{-p_i s } \ell(  x_i,u_i)ds
\end{eqnarray}
where $\gamma_i>0$, $p_i \geq 0$ is a discount factor; here $\ell(x_i,u_i)=-\dot x_i+\gamma_i (u_i-1)$.

A cautionary remark: in the infinite horizon case, the discount factor has the role of ensuring the existence of a finite cost. 
Besides that, looking at \eqref{e2}, we observe that a large value of $p_i>0$ characterizes an "impatient" player who aims at 
fast dissemination of the content. Conversely, a "patient" player would use a small value of $p_i$. 

In particular, we note that  for $p=0$ the cost function has a more familiar expression $J_i(x,u)=-x_i(\tau)+ \gamma_i (\int_0^\tau (u_i(s)-u_{min}) )ds$ where the dependence 
on  the number of copies appears with no discount. In Sec.~\ref{sec:6} we are studying an approximation of our differential game that provides closed form expression of the threshold type for the infinite horizon case for vanishing discounts. In that case, the first term of \eqref{e2} can be approximated assuming very large values of $\tau$ so that 
\[
\int_0^\tau e^{-p_i s }  \dot x_i(s) ds= \frac{ e^{-p_i \tau } x_i(\tau) - z_i}{1 - p_i} \rightarrow  x_i(\tau) - z_i \; \mbox{as} \; p_i  \downarrow 0
\]

Finally, the problem we want to solve is thus to determine the optimal cost function, namely the {\em value function} $V_i(x)$ 
\begin{prob}[Best response]\label{p1} For any strategy $\bu_{-i}$ of the remaining players, determine the best response, i.e., 
the optimal control $u_i^*$ of player $i$ for which the value function is attained, i.e., 
\begin{equation}\label{ep1}
V_i(x):=\inf_{u_i\in \U} J_i(x,u_i)
\end{equation}
\end{prob}

We will solve the problem using the discounted formulation in the infinite horizon: $p_i>0$ for all players, and $\tau=\infty$. This formulation extends to the case of the 
finite horizon either with or without discount and a sketch of the derivation will be provided in Sec.~\ref{sec:6}.

%%%%%%%%%%%%%%%%%%%%%%%%%%%%%%%%%%%%%%%%%%%%%%%%%%%%%%%%%%%%%%%%%%%%%%%%%%%%%%%%%%%%%%%%%%%%%%%%%%%%%%%%%%%%%%%%%%%%%%%%
%%%%%%%%%%%%%%%%%%%%%%%%%%%%%%%%%%%%%%%%%%%%%%%%%%%%%%%%%%%%%%%%%%%%%%

\section{Best response analysis}\label{sec:4}

%%%%%%%%%%%%%%%%%%%%%%%%%%%%%%%%%%%%%%%%%%%%%%%%%%%%%%%%%%%%%%%%%%%%%%
%%%%%%%%%%%%%%%%%%%%%%%%%%%%%%%%%%%%%%%%%%%%%%%%%%%%%%%%%%%%%%%%%%%%%%%%%%%%%%%%%%%%%%%%%%%%%%%%%%%%%%%%%%%%%%%%%%%%%%%%

Best response strategies are determined using the Hamilton-Jacobi-Bellman equation (HJB) for the infinite horizon. %

%%%%%%%%%%%%%%%%%%%%%%%%%%%%%%%%%%%%%%%%%%%%%%%%%%%%%%%%%%%%%%%%%%%%%%
\subsection{Infinite horizon with positive discount $p>0$}
%%%%%%%%%%%%%%%%%%%%%%%%%%%%%%%%%%%%%%%%%%%%%%%%%%%%%%%%%%%%%%%%%%%%%%
The existence of the optimal cost function bounded  and uniformly continuous is immediate from 
\cite{BaCaDo} Prop. 2.8, since indeed the $\ell(\cdot , \cdot)$  is bounded and uniformly Lipschitz, 
since it holds: 
\begin{equation}
|\ell(x,u_i)-\ell(y,u_i)|\leq \lambda_i \umax |x-y|, \ell(x,u_i)|\leq \lambda_i \umax
\end{equation}
In particular, we can write the Hamiltonian for each one of the players with respect to the dynamical system \eqref{e1} 
corresponding to the problem in \eqref{p1}. Before that, it is easy to see that the aggregated dynamics can be 
written as 
\begin{equation}\label{e11}
\dot x = \sum \lambda_i u_i (1 - x)
\end{equation}
which will let us develop the optimal control for each one of the players having 
fixed the control of the competing ones. In particular, the optimal control needs to maximize the Hamiltonian
\begin{eqnarray}\label{e42}
H_i(x,\zeta)&&=\sup_{u \in \U} \left \{-  \sum \lambda_i u_i (1 - x)   \zeta  \right . \nonumber\\
            &&\left . +\big (  u_i\lambda_i (1  - x) -\gamma_i (u_i - u_{min})  \big )\right \}
\end{eqnarray}
Maximization of \eqref{e42} provides the closed loop solution of our problem. However, 
 the optimal cost function $V_i$ in turn is one solving for the HJB equation
\begin{equation}\label{e5}
p_i V_i + H_i \big ( x, \dVx_i \big ) = 0
\end{equation}
so that minimizing the cost function $V_i$ is equivalent to maximize $H_i \big ( x, \dVx_i \big )$ \cite{BaCaDo},  where $\dVx_i=\frac d{dx}V_i(x)$.
In turn, we can write in closed form based on \eqref{e1} and \eqref{e11}.
\begin{eqnarray}\label{e6}
H_i(x)&&\hskip-5mm=u_{min}\gamma_i-\sum_{j\not= i} u_j\lambda_j (1-x)\dVx_i\nonumber\\
         &&\hskip-5mm +\sup_{u_i \in \U} \left \{ u_i \Big [  \lambda_i (1 -x  )\Big (  1 -\dVx_i \Big )  - \gamma_i \Big ] \right \}
\end{eqnarray}
Now, since we observe that the Hamiltonian is linear in $u_i$,  if a maximum is attained at some control $u \in \U$, the intuition 
 is that it may assume only extremal values as it is often seen in the case of open-loop type of solutions \cite{BasarDynGames}. This is actually true: 
 we can allow the control $u$ to take values only on the vertices of the codomain polyhedron, i.e., $[\umin,\umax]^N$. The 
 value function will be the same of the original problem where such restriction does not hold (see \cite{BaCaDo}, pp.113). 
 This is due to the fact that \eqref{e1} is of the type $\dot x:=f_1(x)+f_2(x) u$ and of the running  cost function 
 $\ell$ which is  linear in the control. 
 
 Motivated by this observation, we are interested in a class of best responses, namely
\begin{defi}\label{def:bb}
A strategy $u_i$ is of the bang-bang type if it takes extremal values $\umin$ and $\umax$ only. 
\end{defi}
Also, we denote $t_{ik}, k=1,2,\ldots$  the switching times associated to best response $u_i$, and $[t_{ik},t_{i{k+1}})$ represents 
the corresponding $k$-th switching period of player $i$. If we limit our analysis to the best responses of bang-bang type, then  ${u_i} \in \{\umin,\umax\}$, $i=1,\ldots,N$:
 best responses are in fact piecewise constants. %Also, we denote {\em switching intervals}, the intervals when $\bu$ is constant.

In particular, for the case of bang-bang strategies, the condition for optimality, i.e., a best response, writes from \eqref{e6} 
\begin{equation}\label{e61}
u_i(x)=\begin{cases}
\umin & \mbox{if} \quad   \lambda_i(1-x)\Big (  1 -\dVx_i\Big )  - \gamma_i < 0 \\
\umax & \mbox{if} \quad   \lambda_i(1-x )\Big (  1 -\dVx_i \Big )  - \gamma_i > 0 
\end{cases}
\end{equation}
 We can denote {\em switching interval}
 the interval of time between two consecutive switching instants: within such interval, the best responses 
are constant, i.e., $\bu$ is constant and only assumes values in $\{\umin,\umax\}^N$. We can resume our findings above with the following 
\begin{thm}\label{thm:bb}
The value function $V_i$ corresponding to the best response of the game can be attained by a strategy $u_i$ of the bang-bang type.
\end{thm}
It is worth noting that in general the solution of HJB equations requires to search for a {\em viscosity solution} \cite{BaCaDo}. 
This is due to the fact that the classical solutions assuming the differentiability of the value function may not exist in general and thus require 
to solve for a general notion of differentiation. Here, it is the structure of the system that spares us this step, since we know apriori that the best response  is 
 of the bang-bang type. The main difference with respect to the general case, is that strategies $u_i$s draw values in the finite set $u_i\in \{\umin,\umax\}$; 
 again, compared to the general problem, this fundamental simplification  is due  to the linear structure of the content provider game.

%%%%%%%%%%%%%%%%%%%%%%%%%%%%%%%%%%%%%%%%%%%%%%%%%%%%%%%%%%%%%%%%%%%%%%%%%%%%%%%%%%%%%%%%%
\subsection{Infinite horizon for $p_i>0$}
%%%%%%%%%%%%%%%%%%%%%%%%%%%%%%%%%%%%%%%%%%%%%%%%%%%%%%%%%%%%%%%%%%%%%%%%%%%%%%%%%%%%%%%%%

In order to decide on the sign of the above terms, we need to solve for the HJB equation in $V_i$. This is in 
fact possible, once we notice that \eqref{e5} can be written as the following ODE
\begin{equation}\label{e7}
p_i V_i - a (1-x) DV_i +b_i(1-x)+c=0
\end{equation}
where $V_i$ is the value function that solves for the best response. 

Here we simplify the notation by letting
\begin{eqnarray}\label{e8}
c_i=&&\gamma_i (u_{min}-u_i) \nonumber \\
b_i=&&\lambda_i u_i\nonumber \\
a=&&\sum \lambda_i u_i  
\end{eqnarray}

Whenever convenient, for the sake of clarity, we will denote $a=a(u_i)$ to stress the fact that strategy profile $a$ 
depends on the best response of player $i$, and we will also resort sometimes to the notation 
$a_{-i}:=\sum_{j\not = i} \lambda_i u_i$ the sum of piecewise constant controls played by all the 
remaining players. It is important to note that $a$ and $b_i$ are assumed constant during each
switching interval.

Also, note that since we do not know the optimal control, the expression solving for \eqref{e7} here depends 
on the specific switching interval. Hence, $V_i$ will have a specific dependence on  the control 
which we need to maximize aposteriori since we know that \eqref{ep1} holds.

The solution of the HJB equation above will result can be solved as (see App.~\ref{sec:ODE}) 
\begin{equation}\label{e81}
V_{i}(x)=K\cdot (1-x)^{-\frac {p_i}a} - \frac{p_i b_i(1-x) + (p_i+a)c_i}{p_i(p_i+a)}\\
\end{equation}
where $K$ is a real constant. 

In the following considerations we need the closed form of the function that is maximized by the control $u_i$ in \eqref{e6}
\begin{equation}\label{e812}
\small
\hskip-2mm T_i(u_i,a_{-i})(x)\!=\!u_i \!\left [ \!\Big  ( 1 \!-\! \frac{b_i}{p_i+a} \Big )(1\!-\!x)\! -\! \frac{K p_i}{a}(1-x)^{- \frac{p_i}{a}} \!-\! \frac{\gamma_i}{\lambda_i}  \right ]
\end{equation}

As a first step, from \eqref{e81} we can obtain information on the structure of 
the value function. In particular we resume some basic facts in the following
\begin{lemma}\label{lem1}
i. The best response $u_i^*$ has a finite number of switches.\\
ii. There exists a threshold value of $x_{\infty}$ such that $u_i^*(x)=\umin$ for all $x>x_{\infty}$ for every player $i$.
%iii. Assume $u_i=\umin$ for some $t>0$ and for all players, then no player will ever switch to $\umax$.\\
\end{lemma}
\begin{IEEEproof}
i. By contradiction: assume an infinite number of switches for player $i$ and define constants $K_r$, $r=1,\ldots,\infty$ for each such switching interval. By the continuity of $x$ and 
the continuity of $V_i(x)$, together with \eqref{e81}, there exist an infinite sequence of $K_r$s which are non zero. Hence, since $x\uparrow 1$, by the continuity of $x$, we can
 find sequence $\{x_r\} \uparrow 1$ where $x_r$ belongs to the $r$-th switching interval. Due again to \eqref{e81},  we hence found a subsequence of values of $V_i$ which
  diverges. This is a contradiction since the value function is bounded.  \\
ii. Denote $x_{i,\infty}$ the value of $x$ above which the control switches to $\umin$ for good: indeed the constant appearing in \eqref{e81} is zero. If it was not, again, $V_i$ would grow unbounded as $x\rightarrow 1$. Hence, $DV_i$ is bounded for $x>x_{i,\infty}$, so that by inspection of \eqref{e42}, the control needs to be $\umin$ for values of $x$ close to $1$,  and on the rest of the last switching interval as well since it is constant there.  Finally, we can define $x_{\infty}=\max(x_{i,\infty}, 1\leq i \leq N)$.
\end{IEEEproof}

Furthermore, we can characterize immediately a class of problems where players have no incentive to accelerate anyway: in particular
 we see that the following sufficient condition holds
\begin{lemma}[Degenerate Nash Equilibrium]\label{lem2}
i. Let $\lambda_i < \gamma_i$, then the best response for the player $i$ is $u_i^*=\umin$ irrespective of the other players strategies.
ii. If $\lambda_i < \gamma_i$ for $i=1,\ldots,N$, then $u_i^* =\umin$, $i=1,\ldots,n$  is the unique Nash equilibrium.
\end{lemma}
\begin{IEEEproof}
i.  This is a consequence of the statement in Lemma.\ref{lem1}: in the last switching interval, since $K=0$, from \eqref{e812}
\[
T_i(u_i,a_{-i})(x) \leq \Big ( 1 - \frac{\frac{b_i}{a(\umin)}}{1+\frac{p_i}{a(\umin)}} \Big)- \frac{\gamma_i}{\lambda_i}<0
\]
The rightmost inequality is equivalent to 
\[
\frac{a_{-i}+p_i}{a(\umin)+p_i} < \frac{\gamma_i}{\lambda_i}
\]
which writes also $(\lambda_i - \gamma_i)(a_{-i}+p_i)<\gamma_i \lambda_i \umin$. The statement follows once we observe that the 
previous result is independent of $a_{-i}$, i.e., the strategies played by all the other players.\\
ii. Follows immediately from i.
\end{IEEEproof}

\begin{remark}
From the above results, we can see that the general form of the best response for player $i$ against the strategy profile $\bu_{-i}$ for the remaining players 
can  be determined by proceeding backwards from the latest switching value $x_{i,\infty}$ which is calculated first. Then, by continuity, the constant $K$ appearing in \eqref{e81} for the switching interval 
before the last one (where now player $i$ would use $\umax$) can be determined by imposing the continuity of the value function. In fact, at the switching time, the expression  \eqref{e81} has different values of $b_i$ and $a_i$ in the two adjacent switching intervals. The procedure can be iterated backwards to determine all the threshold values of $x$ when player $i$ switches. 
\end{remark}

In the case of a symmetric game, i.e., when the content providers have all the same parameters, the procedure described above can solve the game in closed form, as showed in 
the next section where symmetric equilibria are described. 

%%%%%%%%%%%%%%%%%%%%%%%%%%%%%%%%%%%%%%%%%%%%%%%%%%%%%%%%%%%%%%%%%%%%%%%%%%%%%%%%%%%%%%%%%%%%%%%%%%%%%%%%%%%%%%%%%%%%%%%%%%%
%%%%%%%%%%%%%%%%%%%%%%%%%%%%%%%%%%%%%%%%%%%%%%%%%%%%%%%%%%%%%%%%%%%%%%%%%%%%%%%%%%%%%%%%%
\section{Symmetric Nash Equilibrium}\label{sec:symplay}
%%%%%%%%%%%%%%%%%%%%%%%%%%%%%%%%%%%%%%%%%%%%%%%%%%%%%%%%%%%%%%%%%%%%%%%%%%%%%%%%%%%%%%%%%
%%%%%%%%%%%%%%%%%%%%%%%%%%%%%%%%%%%%%%%%%%%%%%%%%%%%%%%%%%%%%%%%%%%%%%%%%%%%%%%%%%%%%%%%%%%%%%%%%%%%%%%%%%%%%%%%%%%%%%%%%%%

We consider now the symmetric case: $\lambda_i=\lambda$, $\gamma_i=\gamma$, $p_i=p $ for all $i=1,\ldots,N$. The proofs of the statements hereafter are deferred to the Appendix. Let us consider tagged content provider $i$, and assume that all the remaining players use the same threshold type of strategy, i.e., of the type
\begin{equation}\label{eth}
u_j(x)=
\left \{\begin{array}{ll}
\umax & \text{ if } x< \hat x\\
\umin & \text{ if } x> \hat x \\
\end{array}
\right.
\end{equation}
for some $0\leq \hat x \leq 1$. Denote $x^*$ the last switch of player $i$: we are now ready to show that there exists a symmetric equilibrium where 
also player $i$ will use  $\hat x = x^*$, i.e., the threshold type strategy \eqref{eth} is the best response to itself when all content providers play it. Furthermore, 
it is the unique symmetric equilibrium of the game.

In particular such a Nash equilibrium is given by a threshold $x^*$ 
%\[
%x^*=1-\frac{\gamma(p+n\lambda u_{min})}{\lambda( p+(n-1)u_{min}\lambda)} 
%\]
which is derived by the form of the value function in the last switching interval (recall that $K=0$ in that interval)
\begin{equation}\label{eVlast}
V_{i}(x)= - \frac{ u_{min}\lambda (1-x)}{(p+\fdp{N}\lambda u_{min})}
\end{equation}
by imposing that $T(\umin,\umin(\fdp{N}-1))(x^*)=0$ (switching condition). These results are detailed formally in the statements below. 

%$$
%u_{-i}(x)=
%\left \{\begin{array}{ll}
%u_{max} & \text{ if } x<1-\frac{\gamma(p+n\lambda u_{min})}{\lambda( p+(n-1)u_{min}\lambda)}\\
%u_{min} & \text{ if } x>1-\frac{\gamma(p+n\lambda u_{min})}{\lambda( p+(n-1)u_{min}\lambda)}\\
%\end{array}
%\right.
%$$

%Denote $x^*$ the last switch of the player $i$. Assume that $x^*>1-\frac{\gamma(p+n\lambda u_{min})}{\lambda( p+(n-1)u_{min}\lambda)}$. For $x>x^*$  player's $i$ value function have the following form: 
%$$V_{i}(x)= - \frac{ u_{min}\lambda (1-x)}{(p+n\lambda u_{min})}$$
%
%Indeed because $x^*>1-\frac{\gamma(p+n\lambda u_{min})}{\lambda( p+(n-1)u_{min}\lambda)}$ all players expect $i$ play $u_{min}$. And because of theorem you know that player $i$ plays $u_{min}$. \\

\begin{lemma}\label{lem:switch}
%Let $\lambda(1-\frac{ u_{min}\lambda}{(p+\fdp{N}\lambda u_{min})})-\gamma>0$ and 
\fdp{Let $x^* \geq \hat x$, then the following holds: \\}
i. Player $i$ can switch at some $0<x^*<1$ iff $\lambda(1-\frac{ u_{min}\lambda}{(p+\fdp{N}\lambda u_{min})})-\gamma>0$; moreover  
\begin{equation}\label{esymswitch}
x^*=1-\frac{\gamma(p+\fdp{N}\lambda u_{min})}{\lambda( p+(\fdp{N}-1)u_{min}\lambda)}
\end{equation}
ii. Let all players switch at  $x^*$: the constant $K^*$ which ensures the continuity of the value function $V_i(x)$ at the switching threshold $x^*$ is positive and 
it holds the following relation
\small
%\[
%K=(1-x^*)^{\frac {p}\lambda n u_{max}} \displaystyle{ \frac{p \lambda(1-x^*)( u_{max}(p+n\lambda u_{min})) }{(p+n\lambda u_{min})p(p+\lambda n u_{max})}}
%\]
%\[
% -(1-x^*)^{\frac {p}\lambda n u_{max}}\displaystyle{\frac{ u_{min}(p+\lambda n u_{max})}{(p+n\lambda u_{min})p(p+\lambda n u_{max})}} 
%\]
%\[
% +(1-x^*)^{\frac {p}\lambda n u_{max}}\displaystyle{\frac{ (p+n\lambda u_{min})(p+\lambda n u_{max})\gamma(u_{max}-u_{min}))}{(p+n\lambda u_{min})p(p+\lambda n u_{max})}} 
%\]
%\[
% K^*= (1-x^*)^{\frac {p}\lambda n \umax} \left \{  (1-x^*)  \left [ \frac{1}{n+\frac{p}{\lambda \umax}} - \frac{1}{n+\frac{p}{\lambda \umin}}  \right ] + \frac \gamma{p}(\umax-\umin) \right  \}
%\]
\[
 \!K^*\!=\! (\umax-\umin)\! (1-x^*)^{\frac {p}{\lambda n \umax}}\!\!\left \{\!  \frac{ (1-x^*) \frac{p}{\lambda \umin}}{\Big( \fdp{N}+\frac{p}{\lambda \umax}\Big )\!\!\Big (\fdp{N}+\frac{p}{\lambda \umin}\Big)} \! +\! \frac \gamma{p}\!\right  \}
\]
\end{lemma}

\begin{thm}[Symmetric Nash Equilibrium]\label{thm:nashsym}
Let $\lambda(1-\frac{ u_{min}\lambda}{(p+\fdp{N}\lambda u_{min})})-\gamma>0$, then the threshold type strategy \eqref{eth} where 
$\hat x=x^*$ and $x^*$ is as defined in Lemma~\ref{lem:switch}  is the unique symmetric Nash equilibrium of the game.
%\begin{equation}\label{ens}
%\frac{p}{\lambda}< 2 {n \choose 2} \frac{1-x^*}{K^*} (\umax^*)^2
%\end{equation}   %aa_{-i}\frac{1-x^*}{K^*}$ 
%i. $x_i^*$ defined in Lemma~\ref{lem:switch} is the unique switch for player $i$.
%ii. There exist a unique symmetric Nash equilibrium and the strategy of each player is in threshold form.  
\end{thm}
\begin{IEEEproof}
We first need to ensure that when content provider $i$ plays against \eqref{eth} with $\hat x=x^*$ for all the remaining players, the
switch for player $i$ is unique. Indeed for $x<x^*$ it holds 
\[
D T(u_i,a_{-i})(x)=\frac{- \lambda (1 -\frac{b_i}{a+p})(1-x)^{\frac{p}{a}+1}-\frac{\lambda^2 p K^*}{a^2}}{(1-x)^{\frac{p}{a}+1}}
\]
However, we note that we are in the assumptions of Lemma~\ref{lem:switch}, so that $K^*>0$. 
Thus, for $x<x^*$, indeed $D T(u_i,a_{-i})(x)<0$ by inspection of the above equation. This ensures that 
there is not any other switch for player $i$, so the strategy of player $i$ is also threshold with $\hat x=x^*$.
Indeed,  threshold strategy \eqref{eth} with $\hat x =x^*$ for all players is a best reply to itself for all players, so that it defines
 a Nash equilibrium for the game.  The uniqueness of the equilibrium is obtained by the fact that \eqref{eVlast} has a unique zero.
\end{IEEEproof}

The existence of equilibria in the non symmetric case is the next question that we are 
answering. In particular, we obtain certain asymmetric equilibria which are $\epsilon$-approximated 
Nash equilibria. I.e., the unilateral deviation from those strategy profiles may provide some improvement to the utility 
of a content provider. But, such improvement can be made arbitrarily small by choosing an appropriate value of the discount $p_i$.

%%%%%%%%%%%%%%%%%%%%%%%%%%%%%%%%%%%%%%%%%%%%%%%%%%%%%%%%%%%%%%%%%%%%%%%%%%%%%%%%%%%%%%%%%%%%%%%%%%%%%%%%%%%%%%%%%%%%%%%%%%%
%%%%%%%%%%%%%%%%%%%%%%%%%%%%%%%%%%%%%%%%%%%%%%%%%%%%%%%%%%%%%%%%%%%%%%%%%%%%%%%%%%%%%%%%%
\section{Vanishing discount regime}\label{sec:5}
%%%%%%%%%%%%%%%%%%%%%%%%%%%%%%%%%%%%%%%%%%%%%%%%%%%%%%%%%%%%%%%%%%%%%%%%%%%%%%%%%%%%%%%%%
%%%%%%%%%%%%%%%%%%%%%%%%%%%%%%%%%%%%%%%%%%%%%%%%%%%%%%%%%%%%%%%%%%%%%%%%%%%%%%%%%%%%%%%%%%%%%%%%%%%%%%%%%%%%%%%%%%%%%%%%%%%

Hereafter, we consider the cases of small discount factors. As described in Sec.~\ref{subsec:model}, we can consider the 
case when $p_i$ has a very small value. This means that player $i$ does not pose much of a constraint on the time taken in order to make the content 
popular. In particular, we would consider the case of vanishing discounts sequences: $p_i(r)=o(1)$, $i=1,\ldots,N$ and consider the form 
of the best replies in the regime of vanishing discounts. This provides further insight into the structure of the equilibria for the content providers game.

\begin{cor}\label{thm3}
Let  $p_i(r)=o(1)$: there exists a best reply in threshold form  for the $i$-th player that is arbitrarily close to 
the best reply of the game for a small enough discount factor.
\end{cor}
\begin{IEEEproof}
Let $\zeta_r=\frac{p_i(r)}a$, we  can write \eqref{e812} as 
\begin{eqnarray}\label{e84}
&&\hskip-5mmT(u_i,a_{-i})=u_i \left [ \Big  ( 1 - \frac{\frac{\lambda_i u_i}{ \sum_i  \lambda_i u_i}}{1+\zeta_n} \Big )y - K \zeta_r y^{-\zeta_r} - \frac{\gamma_i}{\lambda_i} \right ] \nonumber \\
&&\hskip-5mm=u_i \left [  \Big ( 1 - \frac{\lambda_i u_i}{\sum_i  \lambda_i u_i}\Big ) (1 - \zeta_r + o(\zeta_r)) y - \frac{\gamma_i}{\lambda_i}  \right . +\\
&& \left . K \zeta_r (y u_i - \zeta_n + o(\zeta_r))\Big )\right ] \nonumber \\
&&\hskip-5mm=u_i  \Big [ \Big ( 1 - \frac{\lambda_i u_i}{\sum_i  \lambda_i u_i}\Big )y - \frac{\gamma_i}{\lambda_i} \Big  ]+\zeta_r  f(y) + o(\zeta_r) \nonumber
\end{eqnarray}
where $y=1-x$ and  $f(y)=u_i \big ( y\frac{\lambda_i u_i}{\sum_i  \lambda_i u_i} - 1/y^{\zeta_r}\big )$. 
We already noticed that there exists $x_\infty$ above which every player switches to $\umin$: we can hence restrict our discussion to 
the range $y \in [1-x_\infty, 1]$. Indeed, $f(y)$ is bounded therein: denote $\tilde T(u_i,a_{-i}) =u_i \Big( \frac{\lambda_i u_i}{\sum_i  \lambda_i u_i} - \frac{\gamma_i}{\lambda_i}  
\Big )$.  

Hence, we can fix $\epsilon>0$ and consider $r>r_o$ such that $|\tilde T(u_i) - T(u_i)|<\epsilon$ uniformly in $y$.  The best response of the user $i$ will at most 
produce a value function that differs by $\epsilon$ from the one which maximizes \eqref{e6}. Hence, we search for the solution of the maximization problem 
\[
\tilde u_i^*=\argmax\limits_{u_i\in\{\umin,\umax\}} \tilde T(u_i)
\]
which corresponds to a modified game where the cost function is $\tilde T$. We hence need to state when $\tilde T(\umin,a_{-i})$ is larger or smaller than $\tilde T(\umax,a_{-i})$: this turns out to be equivalent to the condition for the state $x$ to exceed or not the threshold
\begin{eqnarray}\label{e85}
%x_{0,i} := 1 - \frac{\gamma_i}{\lambda_i}\Big (1 + \frac1{\frac{\umax+\umin}{a_{-i}/\lambda_i}+\frac{\umax\umin}{a_{-i}^2/\lambda_i^2}} \Big )
x_{0,i} := 1 - \frac{\gamma_i}{\lambda_i} \frac{1}{\big ( 1+\frac{\lambda_i \umin}{a_{-i}} \big ) \big ( 1+\frac{\lambda_i \umax}{a_{-i}} \big )}
\end{eqnarray}
so that the final control law that governs the best reply of the $i$-th player. is 
\begin{equation}\label{e62}
\tilde u_i^*(x)=\begin{cases}
\umax & \mbox{if} \quad   x \leq x_{0,i} \\
\umin & \mbox{if} \quad    x > x _{0,i}
\end{cases}
\end{equation}
which concludes the proof. 
\end{IEEEproof}
\begin{remark}
Because of the above result, we can always find a discount factor small enough so as to find a threshold type strategy which approximates 
the cost function of the best response within an arbitrarily small positive additive constant $\epsilon>0$. In turn, this also means that if a Nash equilibrium 
exists under the modified utility function $\tilde T(\cdot)$, then it is  an $\epsilon$-approximated Nash equilibrium in threshold policy for 
the original game. It is hence interesting to study the existence of a Nash equilibrium for the modified game. 
\end{remark}
\begin{lemma}\label{thm4}
Consider $\tilde u_i^*(t)$ in the modified game and player $i$ switching at $t_k$: % either $u_i^*(t_k^-)=\umax$ $u_i^*(t_k^+)=\umin$, i.e., 
if  $i$ switches is from $\umin$ to $ \umax$ then some other player switches from $\umax$ to $\umin$ at $t_k$.
\end{lemma}
\begin{IEEEproof}
By contradiction, assume that there exist a single player such that switches from $\umin$ to $\umax$ at switching time 
$t_k$: from \eqref{e85}, it is clear that $x_{0,i}$ do not change passing from the $k$-th switching interval to the new one
 because $a_{-i}$ is unchanged. But, this means that $x(t_k^+)<x_{0,i}<x(t_k^-)$. Of course, this is not possible since the 
 dynamics of $x$ is monotone non decreasing. In the same manner, it is easy to see that if only switches occur from $\umin$ 
 to $\umax$, $a_{-i}(t_k˜^-) > a_{-i}(t_k^+)$ so again $x(t_k^+)<x_{0,i}<x(t_k^-)$.
\end{IEEEproof}

\begin{thm}[Asymmetric $\epsilon$- approximated Nash Equilibrium]\label{thm5}
Let $\lambda_i=\lambda$ for $i=1,\ldots,N$ and $\lambda>\gamma_1 > \gamma_2 > \ldots > \gamma_N$.
 Then, there exists an $\epsilon$-approximated Nash equilibrium in threshold form for the original game.
\end{thm}
\begin{IEEEproof}
We assume the regime of vanishing discounts such in a way that $\epsilon$ is defined in the sense of Thm.~\ref{thm3}. 
The proof is based on the following observation: at time $0$, indeed $u_i^*(0)=\umax$ because of \eqref{e85} and $\lambda>\gamma_i$ for 
all $i$s. Clearly, since $\gamma_1> \gamma_i$ for $i>1$, then $x_{0,1}<x_{0,i}$ for $i>1$, and $t_1$ corresponds to the switch of node $1$ from $\umax$ to $\umin$. 
Also, ${a_{-i}(t_1^-)}=\umax N> \umax (N-1) + \umin = a_{-i}(t_1^+)$ for all players  $i>1$, so that $x_{0,i}(t_1^-)<x_{0,i}(t_1^+)$. Thus, 
$u_i^*(t_1^-)=u_i^*(t_1^+)=\umax$. Finally, until $x<x_{0,2}(a_{-i}(t_1^+))$, all players not $i$ will use $\umax$.
 
 By induction: assume that first $k-1$ players that switched from $\umax$ to $\umin$ did not switch back and prove that under the 
conditions in the assumptions even the $k$-th player will never switch from $\umin$ to $\umax$. %First, observe that this event could only be 
%possible Thm~\ref{thm4}

In order to proceed further with the proof we need to precise some notation 
\begin{itemize}
\item $x_{0,i}(k)$ is the threshold \eqref{e85} for player $i$ when $k$ players already switched to $\umin$;
\item $a_{-i}(k)$ is the sum of the other players $\lambda_j u_j$ when $k$ of them switched to $\umin$.
\end{itemize}
At the time when player $k+1$ switches, it holds $x(t_{k+1})=x_{0,k}(k)$. Hence, in order for player $k+1$ 
not to switch back to $\umax$, it must hold $x(t_{k+1})>x_{0,k}(k+1)$.  However, we know that the dynamics 
in the $k$-th switching period is 
\[
x(t_{k+1})=1 - (1-x_{0,k}(k-1)))e^{-a_k(t_{k+1}-t_{k})}
\]
Also, by inductive assumption, $x_{0,k}(k-1)=x_{0,k}(k)$ because no player switched back to $\umax$. Then, since $a_k$ 
is constant in the $k$-th switching period, the dynamics in that interval is governed by
condition  
\begin{equation}\label{e51}
\frac{1-x_{0,k}(k+1)}{1-x_{0,k}(k)}>e^{-a_k\Delta t_{k+1}} % (t_{k+1}-t_k)}
\end{equation}
Moreover, we have a condition on $\Delta t_{k+1}=t_{k+1}-t_k$:
\begin{eqnarray}\label{e52}
x_{0,k}(k)=1-(1-x_{0,k}(k))e^{-a_k (t_{k+1}-t_k)} \nonumber \\
\Rightarrow \Delta t_{k+1} = \log \Big ( \frac{1-x_{0,k+1}(k)}{1-x_{0,k}(k)}\Big )
\end{eqnarray}
Now, combining \eqref{e51} and \eqref{e52} we obtain 
\begin{equation}\label{e53}
\frac{1-x_{0,k}(k+1)}{1-x_{0,k}(k)}> \frac{1-x_{0,k+1}(k)}{1-x_{0,k}(k)} \Rightarrow x_{0,k}(k+1) < x_{0,k+1}(k)
\end{equation}
We can now express the condition above by considering the explicit expression 
%\[
% x_{0,k}(k+1)=1-\frac{\gamma_k}{\lambda_k} \Big ( 1+\frac{v_k^2}{v_k(\umax+\umin)+\umin\umax}\Big )
%\]
\[
 x_{0,k}(k+1)=\frac{\gamma_k}{\lambda} \frac{1}{\big ( 1+\frac{ \umin}{v_k} \big )\big ( 1+\frac{ \umax}{v_k} \big )}
\]
where $v_k:=\frac{a_{-k}(k+1)}{\lambda_k} $. Also, in the same way, 
%\[
% x_{0,k}(k+1)=1-\frac{\gamma_k}{\lambda_k}  \Big ( 1+\frac{v_{k+1}^2}{v_{k+1}(\umax+\umin)+\umin\umax}\Big )
%\]
\[
 x_{0,k+1}(k)=\frac{\gamma_k}{\lambda} \frac{1}{\big ( 1+\frac{ \umin}{v_{k+1}} \big )\big ( 1+\frac{ \umax}{v_{k+1}} \big )}
\]
where $v_{k+1}:=\frac{a_{-(k+1)}(k)}{\lambda_k} $. Finally, let us observe that 
\[
v_k=(N-1-k)\umax+k \umin = v_{k+1}
\]
so the condition in \eqref{e53} becomes $\gamma_k>\gamma_{k+1}$ which is 
true according to our assumptions. Hence the inductive step is complete and the 
statement is true. 
\end{IEEEproof}
\begin{remark}
We note  that the constructive proof of the $\epsilon$-approximated Nash equilibrium confirms the following intuition: if some player 
does not accelerate any longer, it will not have incentives to accelerate later for larger values of the state $x$, because the increment in the 
state $x$ is decreasing. Overall, the above statement suggests that in the fully asymmetric case, the presence of diverse costs induces 
an equilibrium in threshold form where even if a content provider has an incentive in deviating from the given strategy profile, and so change 
strategy, the incentive that the player has in deviating can be made small at wish by choosing an appropriate value of the discount.
\end{remark}

In the next section we sketch how the framework proposed for the infinite horizon can be extended in the case of a finite horizon. 

%%%%%%%%%%%%%%%%%%%%%%%%%%%%%%%%%%%%%%%%%%%%%%%%%%%%%%%%%%%%%%%%%%%%%%%%%%%%%%%%%%%%%%%%%%%%%%%%%%%%%%%%%%%%%%%%%%%%%%%%%%
%%%%%%%%%%%%%%%%%%%%%%%%%%%%%%%%%%%%%%%%%%%%%%%%%%%%%%%%%%%%%%%%%%%%%%%%%%%%%%%%%%%%%%%%%
\section{Finite horizon case}\label{sec:6}
%%%%%%%%%%%%%%%%%%%%%%%%%%%%%%%%%%%%%%%%%%%%%%%%%%%%%%%%%%%%%%%%%%%%%%%%%%%%%%%%%%%%%%%%%
%%%%%%%%%%%%%%%%%%%%%%%%%%%%%%%%%%%%%%%%%%%%%%%%%%%%%%%%%%%%%%%%%%%%%%%%%%%%%%%%%%%%%%%%%%%%%%%%%%%%%%%%%%%%%%%%%%%%%%%%%%

When there is a finite horizon $0\leq \tau < \infty$ under a nonnegative discount, the HJB equation becomes \cite{BaCaDo}
\begin{equation}\label{e9}
\dot V_{i} + p V_i  + H \big ( x, DV_{i}  \big ) = 0
\end{equation}
The natural initial condition $V_i(x,0)=0$ for all $x\in \R$, because the terminal cost is null. 
Hence, the value function solves the following PDE  
\begin{equation}\label{e92}
\dot V_{i} + p V_i  - a (1-x) DV_i +b(1-x)+c=0
\end{equation}
\eqref{e92} is linear and the associated homogeneous PDE is 
\begin{equation}\label{e10}
\dot V_{i} + p V_i  - a (1-x) DV_i = 0 
\end{equation}
whose solution is in the form $V_i^{om}(x,t)=\phi(a^{-1}\log(1-x)-t)(1-x)^{-\frac pa}$, where $\phi(v):\R \rightarrow \R$ 
is a differentiable function.

Hence we just need a particular solution: we seek one such 
solution in the form $V_i(x,t)=V_{i,p}(x)$, so that it should solve 
$$p_i V_i  - a (1-x) DV_i +b(1-x)+c=0$$
The solution is found to be: 
\begin{equation}\label{e12}
V_{i}(x)=\begin{cases}
&(1-x)^{-\frac {p_i}a} - \frac{p_i b(1-x) + (p_i+a)c}{p_i(p_i+a)} \quad \mbox{if} \; p_i>0\\
&\frac ba x - \frac ca \log (1-x) \quad \mbox{if} \; p_i=0
\end{cases}
\end{equation}

%$$ V_p(x) = c + \frac{1}{(e-p)\,e\,p} \, {\left({\left(p \, e^{\left(\frac{e-p}{e} \,\log\left(f + e \, x\right)\right)} + (e-p) \, f e^{\left(-\frac{p}{e} \,\log\left(f + e \, x\right)\right)}\right)} g - (e-p)e \, h \, e^{\left(-\frac{p}{e} \, \log\left(f + e \, x\right)\right)} \right)} e^{\left(\frac{p}{e} \, \log\left(f + e \,\right)\right)}$$
Finally, the solution of \eqref{e92} is determined to be 
\[
V_i(x,t)=\phi(a^{-1}\log(1-x)-t)(1-x)^{-\frac {p_i}a} + V_{i,p}(x) 
\]

Since we are faced with an undetermined function $\phi$, one per switching interval, we shortly describe 
how to calculate the best response. In the first switching interval $[0,t_1]$, the natural initial condition $V(x,0)=0$ for all $x \in [0,1)$.
This provides the closed form expression for $\phi(\cdot)$, which is found by 
\[
\phi(a^{-1}\log(1-x)-t) (1-x)^{-\frac pa}+V_p(x)=0, \quad \forall x \in \R.
\]
%In the case $p=0$, this can be written as 
%\[
%\phi(y)=-\frac ba(N - e^{ay} ) + c y
%\]
%so that finally we obtain 
%\begin{equation}\label{e13}
%V(x,t) =-ct+\frac ba \Big [ (N-x)^{1/a}e^{-at } - (N-x) \Big ]
%\end{equation}
%
%Using the same condition, we can solve for the case $p>0$. 
In particular, we obtain for the case $p>0$:
\[
\phi(v)=-1+e^{pv}\Big ( \frac{b}{p+a}e^{av} +\frac cp \Big )
\]
so that in the first switching interval  we can state
\begin{equation}\label{e14}
V(x,t) =-\Big ( 1 - e^{-pt}\Big ) \Big [ \frac cp  + \frac b{p+a}(1-x) \Big ]
\end{equation}
Now, once determined the best response for player $i$ in the first switching interval by \eqref{e6}, we should impose the continuity condition on $V(x,t)=V(x,t_1)$. 
This provides the initial condition for the second interval. Proceeding to the subsequent intervals, the procedure can be iterated to determine 
 the value function for the best response of player $i$. It is worth noting that in this case the switching thresholds will depend on time. 
%%%%%%%%%%%%%%%%%%%%%%%%%%%%%%%%%%%%%%%%%%%%%%%%%%%%%%%%%%%%%%%%%%%%%%%%%%%%%%%%%%%%%%%%%%%%%%%%%%%%%%%%%%%%%%%%%%%%%%%%
\begin{figure*}[t]
  \centering 
        \captionsetup{belowskip=-15pt,aboveskip=-10pt}
        \captionsetup[subfigure]{labelformat=empty}
          \subfigure[]{\includegraphics[width=0.24\textwidth]{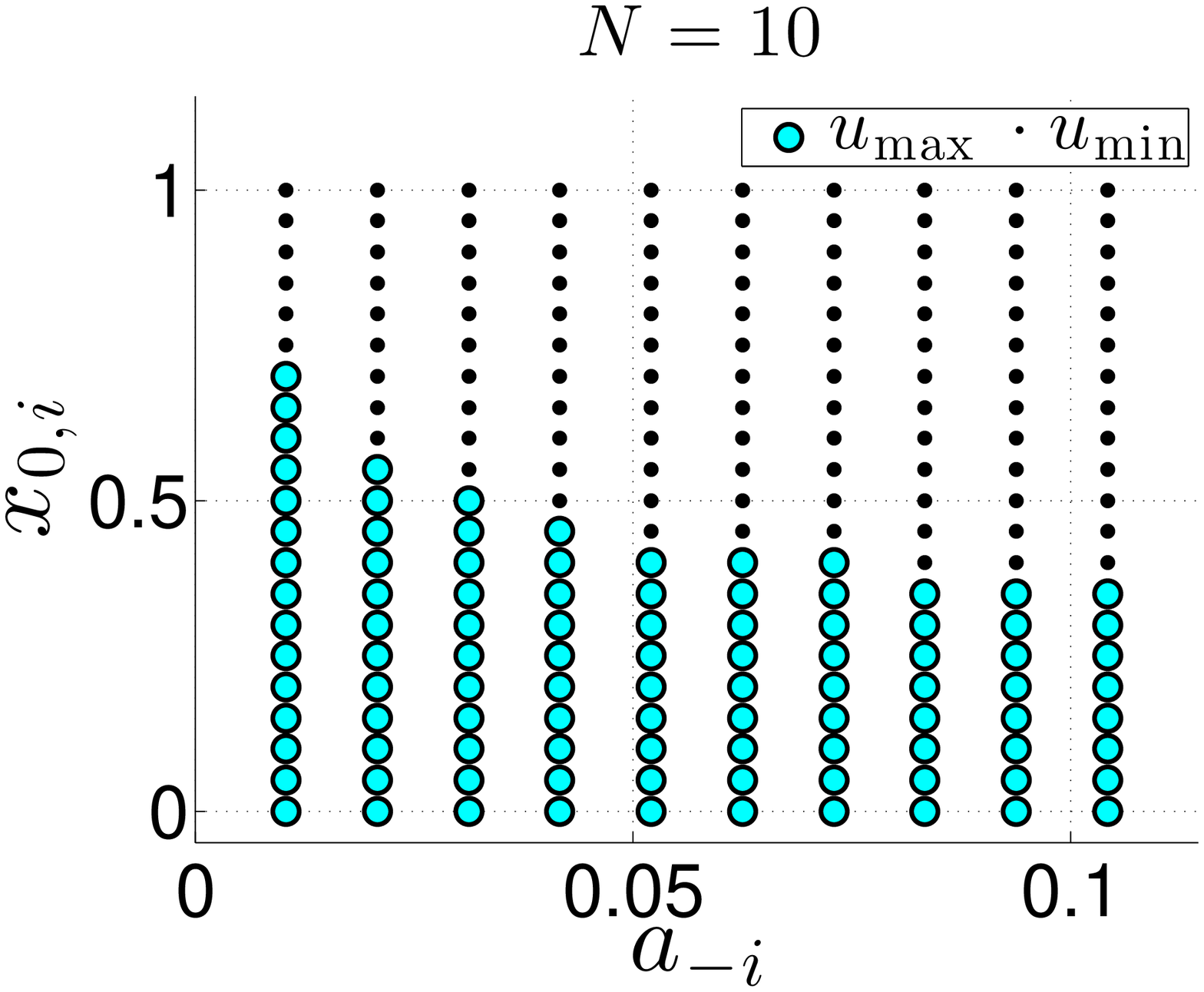}  \put(-121,95){a)} }
          \subfigure[]{\includegraphics[width=0.24\textwidth]{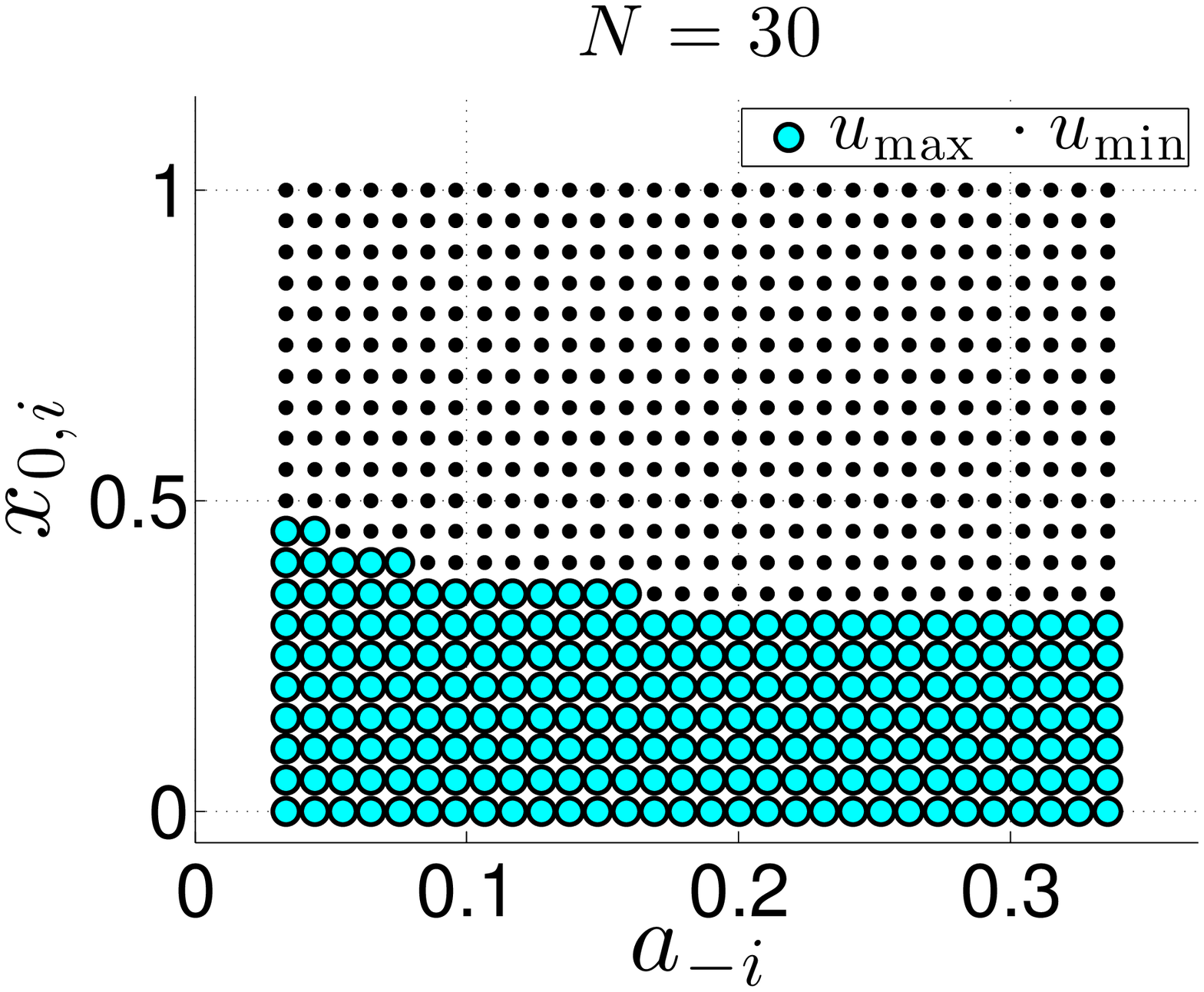} \put(-121,95){b)}  }
         \subfigure[]{\includegraphics[width=0.24\textwidth]{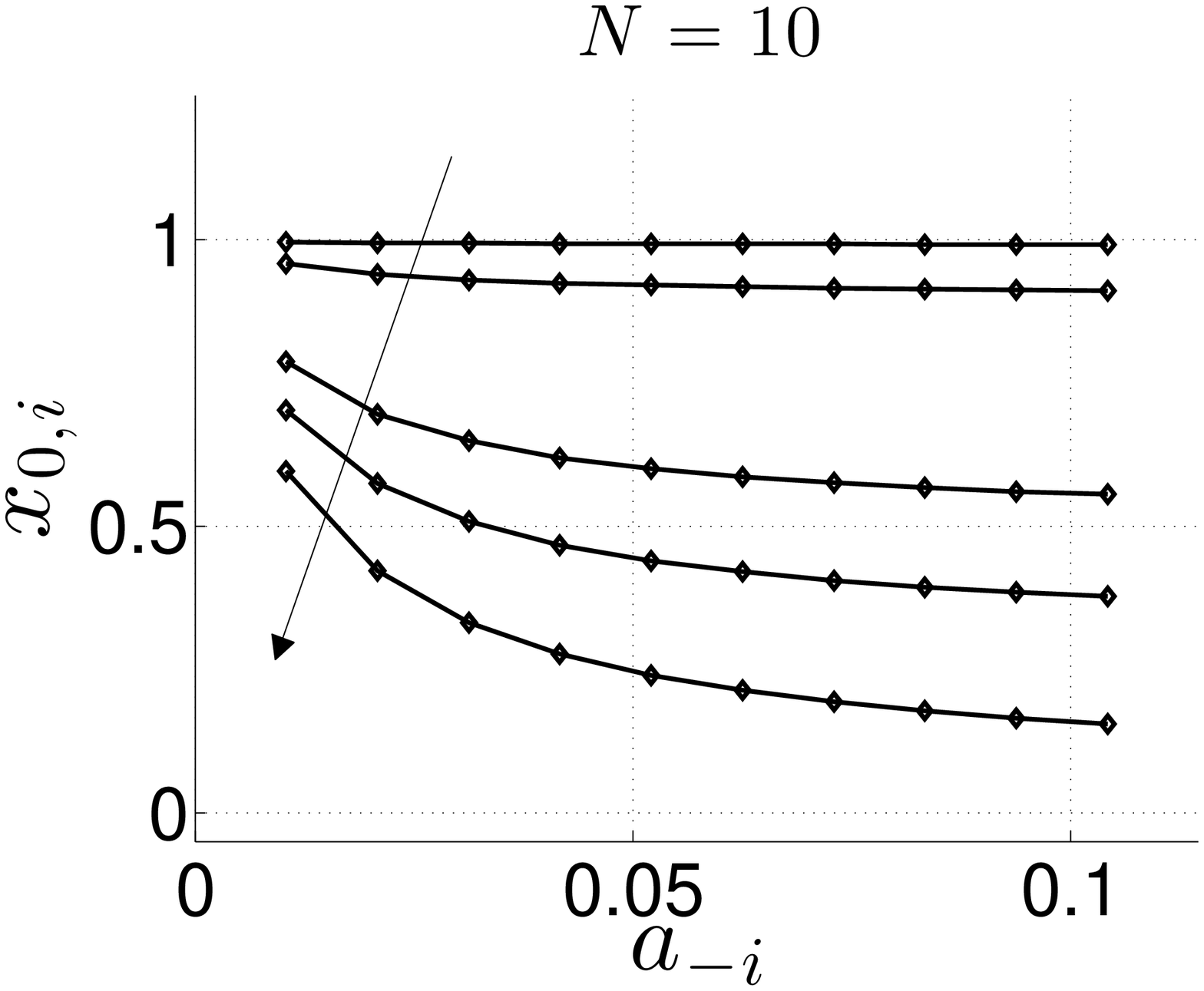} \put(-121,95){c)} \put(-88,87){\scriptsize $\frac \gamma\lambda=0.01,0.1,0.5,0.7,0.95$}  }
          \subfigure[]{\includegraphics[width=0.24\textwidth]{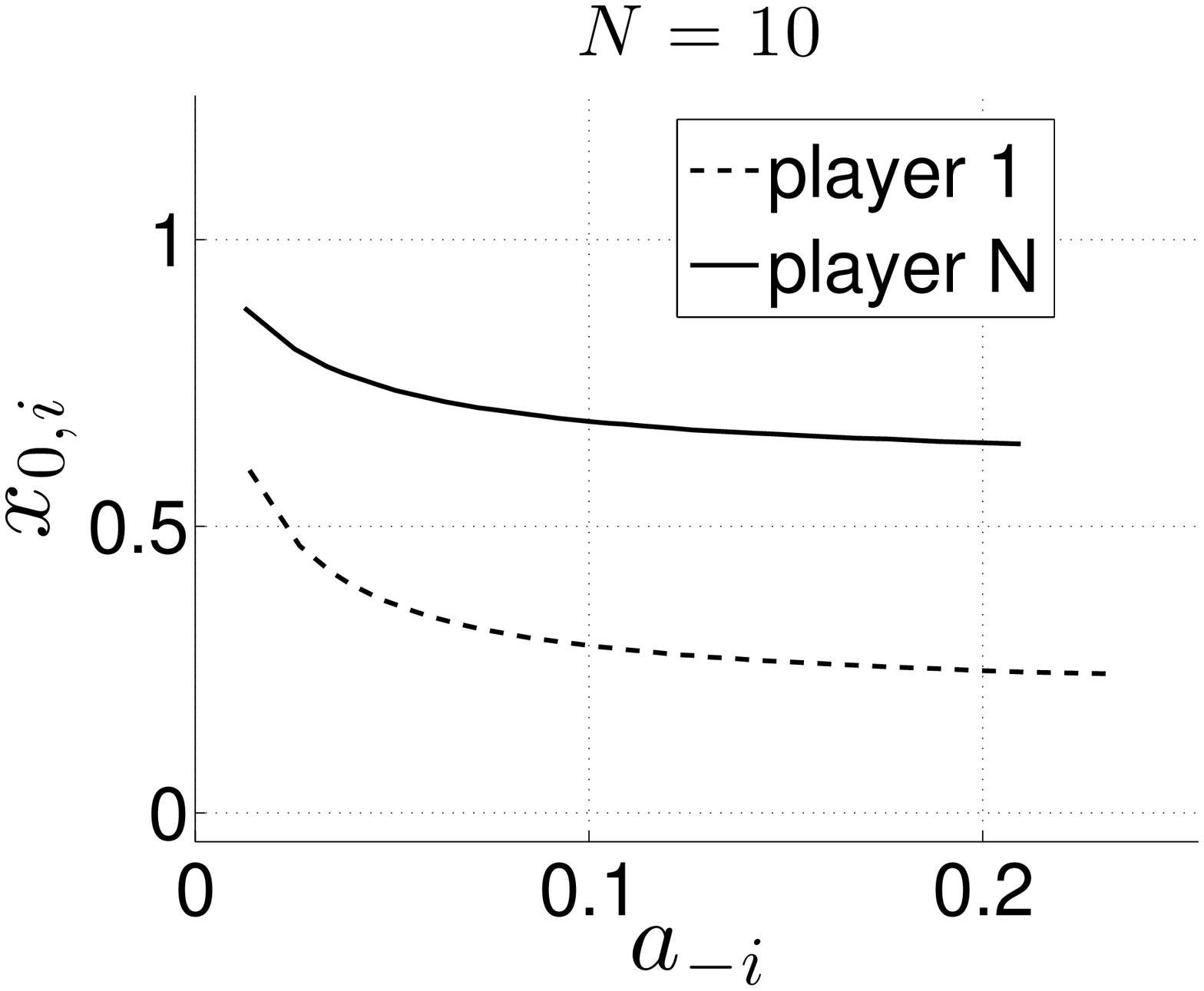}\put(-121,95){d)} }
%      \put(-210,120){$X(t,\theta)$}
%      \put(-110,10){$t_\beta(\theta)$}\put(-30,10){$\tau$}\put(-80,5){$\tau-t_\beta(\theta)$}  
%      \put(-120,110){$\beta$}      
  \caption{a) and b): best reply in the case $\lambda_i=\lambda_0=100$ views/day, $\umax=10$, $\umin=1$, $p=\lambda$, and $\gamma_i=\gamma=0.7 \lambda$; a) $N=10$ and b) $N=30$; c): impact of the cost on the best response; d) case for heterogeneous scenario: $\lambda_i=\lambda_0$ for $i=1,\ldots,N/2$ and $\lambda_i=2\lambda_0$ for 
  $i=N/2+1,\ldots,N$, $N=10$.}\label{fig:accel}
\end{figure*}

%%%%%%%%%%%%%%%%%%%%%%%%%%%%%%%%%%%%%%%%%%%%%%%%%%%%%%%%%%%%%%%%%%%%%%%%%%%%%%%%%%%%%%%%%%%%%%%%%%%%%%%%%%%%%%%%%%%%%%%%
%%%%%%%%%%%%%%%%%%%%%%%%%%%%%%%%%%%%%%%%%%%%%%%%%%%%%%%%%%%%%%%%%%%%%%
%%%%%%%%%%%%%%%%%%%%%%%%%%%%%%%%%%%%%%%%%%%%%%%%%%%%%%%%%%%%%%%%%%%%%%%%%%%%%%%%%%%%%%%%%%%%%%%%%%%%%%%%%%%%%%%%%%%%%%%%
%%%%%%%%%%%%%%%%%%%%%%%%%%%%%%%%%%%%%%%%%%%%%%%%%%%%%%%%%%%%%%%%%%%%%%

\section{Numerical Results}\label{sec:7}

%%%%%%%%%%%%%%%%%%%%%%%%%%%%%%%%%%%%%%%%%%%%%%%%%%%%%%%%%%%%%%%%%%%%%%
%%%%%%%%%%%%%%%%%%%%%%%%%%%%%%%%%%%%%%%%%%%%%%%%%%%%%%%%%%%%%%%%%%%%%%%%%%%%%%%%%%%%%%%%%%%%%%%%%%%%%%%%%%%%%%%%%%%%%%%%

In this section we provide a numerical description of the results for the best responses of content providers.  
It is interesting to visualize the best reply of a certain content provider facing different strategies 
of the remaining ones. In Fig.~\ref{fig:accel}a) and b) we reported on the best response in the 
vanishing discount regime as a function of $a_{-i}$ in the homogeneous scenario, i.e., 
$\gamma_i=\gamma$ and $\lambda_i=\lambda_0$ for $i=1,\ldots,N$. Here, $\umax=10$ and $\umin=1$.

The graphs of Fig.~\ref{fig:accel}a) and b) refer to two values $N=10$ and $N=30$, respectively. 
The value of $\lambda_0$ is settled to $100$ views per day, whereas  $\gamma=0.7\cdot \lambda$. 

The best response is depicted using different markers for $\umin$ and $\umax$: for a fixed value 
of $a_{-i}$ the best response starts with $\umax$ and switches to $\umin$ above the threshold 
$x_{0,i}$. As it can be noticed in both cases a) and b)  the threshold value decreases with $a_{-i}$. 
This is the  effect of competition: larger values of $a_{-i}$ correspond to more players using $\umax$. 
Hence, the values of $x$ when the residual number of views that can be expected for a certain content 
are too small and the cost for accelerating takes over. Hence the switch occurs at lower values of $x$. 

Finally, we observe that there is a floor at $1-\frac{\gamma}{\lambda}$: this is the limit 
best response for large number of players as from \eqref{e85}. In the case of many players in the game, 
i.e., very large number of contents,  the best response of the single player should depend on $\gamma$ and $\lambda$ only. When this 
happens, in fact, the game becomes singular, and there exists a unique best response for every player, which defines the only equilibrium of the system.\footnote{Strictly speaking,
  in the limit of large number of players, the action of a content provider become independent of the actions of the others.} 

In Fig.~\ref{fig:accel}c) we observe the best response for $N=10$ and for increasing  values of  
$\gamma/\lambda$: clearly, larger values of the ratio makes players accelerate less because of  
increased cost. Again, we observe that the effect of competition is to reduce the acceleration for 
larger values of players using $\umax$, i.e., larger values of $a_{-i}$. However, 
for small values of the cost, not only the best reply is to use a large threshold, but the strategy of competing 
content providers becomes less and less relevant so that the threshold becomes almost constant in $a_{-i}$.

In the last figure, i.e., Fig.~\ref{fig:accel}d), we considered an heterogeneous scenario. In this case, 
contents $i=1,\ldots,5$ have $\lambda_i=\lambda_0$, whereas $i=6,\ldots,10$ have $\lambda_i=2\lambda_0$.
In this case we expect that  $a_{-1}$ and $a_{-10}$  in the example may be different. Hence, we are interested
in the relative behavior of best response of the two type of players. 

As seen in the figure,  the switching order $x_{0,1}<x_{0,10}$ is maintained for increasing values
 of $a_{-i}$: players with higher value of $\lambda_i$ always switch before. Actually, even for a small 
 number of players as in this example,  the range of values taken by $a_{-i}$ is basically the same for both 
  types of players. As a consequence, the ratio $\gamma_i/\lambda_i$ is the main parameter 
  characterizing the relative behavior of the two classes of players, i.e., the switching order.

%%%%%%%%%%%%%%%%%%%%%%%%%%%%%%%%%%%%%%%%%%%%%%%%%%%%%%%%%%%%%%%%%%%%%%%%%%%%%%%%%%%%%%%%%%%%%%%%%%%%%%%%%%%%%%%%%%%%%%%%
%%%%%%%%%%%%%%%%%%%%%%%%%%%%%%%%%%%%%%%%%%%%%%%%%%%%%%%%%%%%%%%%%%%%%%

\section{Conclusions}\label{sec:9}

%%%%%%%%%%%%%%%%%%%%%%%%%%%%%%%%%%%%%%%%%%%%%%%%%%%%%%%%%%%%%%%%%%%%%%
%%%%%%%%%%%%%%%%%%%%%%%%%%%%%%%%%%%%%%%%%%%%%%%%%%%%%%%%%%%%%%%%%%%%%%%%%%%%%%%%%%%%%%%%%%%%%%%%%%%%%%%%%%%%%%%%%%%%%%%%

In this paper we have introduced models for advertisement in online content diffusion. 
In this context, the key observation is that the competition in order to make contents popular defines
 a dynamic game among content providers. They leverage on acceleration tools of online platforms in order to 
increase the viewcount of their contents. However, there is a fee to pay in order to profit from the re-ranking  
of recommendation lists and become featured, i.e., to occupy positions that are more visible on the web pages of 
potential viewers.

As such, each player, e.g., each content provider, needs to decide over time if it is worth to accelerate or not. And, this 
choice is dynamic over time and it depends on what other content providers do, since the viewers' base is shared. We 
leveraged on the framework of differential games. In differential games,  the best response of single players is determined 
by solving an ODE involving the Hamiltonian of the system observed by a single player. 
 
 We showed that in the infinite horizon case, the closed loop best replies of players are of the bang-bang type in the 
 state $x$. Much of the machinery involved in our proofs was made possible  by the specific structure of the problem. Thus, 
 we were able to identify a unique threshold-type Nash equilibrium in the symmetric case and we found that dual counterparts 
 exist  in the fully asymmetric case for small discounts.

 \fdp{{\em Practical implications.} We would like to highlight some practical implications which can be derived 
 from our model. First we notice that we have been able so far to derive the existence of Nash equilibria in threshold form 
in the symmetric and ($\epsilon$-approximate) in the asymmetric case. However, we conjecture that Nash 
equilibria in threshold form do exist also for any choice of the $\lambda_i$s and the $\gamma_i$s. 
Thus, content providers would only pay when the total fraction of views is below a certain 
threshold and stop promoting above it. Now, we can observe \eqref{esymswitch} closer, and draw the 
following conclusions when the equilibrium is reached:\\}
\fdp{a) {\em all-or-nothing effect}: a content $i$ with low potential, i.e., 
very small $\lambda_i$ will not be accelerated at all when $\lambda_i < \gamma_i$. This suggests 
that a content provider should always compete by promoting first contents that are likely to 
become most popular even without promotion, e.g., those with larger values of $\lambda_i$.\\}
\fdp{b) {\em best response and promotion}: the best response is of the threshold type, so
that content providers are able to maximize the number of views while minimizing the cost 
by using the maximum promotion budget per day they have until the threshold is reached. 
This means that, from the content provider perspective, the acceleration tools available in practice, 
such as the YouTube promotion campaign, can well be used to optimize for the tradeoff between 
costs and acceleration.\\}
\fdp{c) {\em daily budget:} the daily budget $\gamma$ determines the threshold $x^*$, such in a way that 
the larger the cost which is paid per day, the lower $x^*$. Now let us take the platform owner perspective:
for a given cumulative budget paid by a customer, the smaller the threshold, the lesser the promotion time will last. 
This is indeed  the better option for the sake of system's resources; hence,  larger acceleration fares will 
lead to shorter  promotion campaigns with indeed lesser load for the platform promotion mechanisms. } \\
 {\em Future works.} The results of our paper indicate that these game models can lead to new tools for the 
 pricing of online content advertisement and for the prediction of content popularity. To this respect, this work is 
 by no means conclusive since there are several interesting research directions that are left for future work. First, the dynamic setting 
 in the finite horizon case appears the most immediate extension. We have showed that the value function of each player can be derived in closed form. 
 However, we have not been yet investigating the structure of the equilibria for that game. In future work, we plan to study the effect of the 
 horizon duration onto the equilibria and the effect that time constraints have on content providers' strategies. Another aspect which was left out of the scope of this work relates to the number of competitors: in the case of large $N$, the strategy of single players does not change significantly other players' utility and the strategy profile   $\bu_{-i}$. To this respect, the dynamic game formulation could be reduced in the limit of large $N$  to a static formulation which  could be studied using Wardrop-like  equilibria \cite{NetSciCom}. 
%%%%%%%%%%%%%%%%%%%%%%%%%%%%%%%%%%%%%%%%%%%%%%%%%%%%%%%%%%%%%%%%%%%%%%%%%%%%%%%%%%%%%%%%%%%%%%%%%%%%%%%%%%%%%%%%%%%%%%%%
%%%%%%%%%%%%%%%%%%%%%%%%%%%%%%%%%%%%%%%%%%%%%%%%%%%%%%%%%%%%%%%%%%%%%%
\appendix
\subsection*{ODE solution}\label{sec:ODE}

The solution of \eqref{e7} is equivalent to the solution of $DV_i -\frac{p_i}{a(1-x)} V_i=\frac{b_i}a+ \frac{c}{a(1-x)} $, so that it 
is sufficient to observe that the integrating factor for this first order ODE is $(1-x)^\frac{p_i}{a}$. Hence the solution follows from
\begin{eqnarray}
&&\hskip -3mm V_i(x)=(1-x)^{-\frac{p_i}{a}} a^{-1}\int (1-x)^\frac{p_i}{a}\Big ( b_i + \frac{c_i}{1-x}\Big )dx \no\\
        &&\hskip -3mm =(1-x)^{-\frac{p_i}{a}} \Big [ -\frac{b_i}{a+p_i}(1-x)^{1+\frac{p_i}{a}}-\frac c{p_i} (1-x)^\frac{p_i}{a} + K \Big ] \no \\
        &&\hskip -3mm =K (1-x)^{-\frac{p_i}{a}} - \frac{b_i}{a + p_i} (1-x) - \frac c{p_i}
\end{eqnarray}
where $K$ is an arbitrary real constant.

\subsection*{Proof of Lemma~\ref{lem:switch}}
\begin{IEEEproof}
(i) %Let $T(x)=\lambda(1-x)(1-DV_i)-\gamma$. 
The proof is made in three steps. In the first step we prove that $T(\umin,(N-1)\umin)(x)$ is decreasing for $x>x^*$ (we use 
notation $T(x)$ when it does not generate confusion). In the second step we derive the sufficient condition in the assumptions.
Finally, in the third step we compute $x^*$ and the corresponding constant $K^*$.

{\em  Step 1.} From  \eqref{eVlast}, in the last switching interval we have
$$DV_i=  \frac{ \umin\lambda}{(p+\fdp{N}\lambda \umin)}$$
If we plug in $DV_i$ in $T(x)$ we finally have:
$$
T(x)=(1-x)(1-\frac{\umin\lambda}{(p+\fdp{N}\lambda \umin)})-\frac{\gamma}{\lambda}
$$ 
so that $D T(x)<0$.

{\em Step 2.} Since $T(x)$ is decreasing in the last switching interval, a threshold when player $i$ switches to $\umin$ exists if and only if $T(0)>0$, which  is the assumption in the statement, namely $1-\frac{ \umin\lambda}{p+\fdp{N}\lambda \umin } >\frac{\gamma}{\lambda}$ . 

{\em Step 3.} The threshold $x^*$ for player $i$ is obtained by solving 
$$
T(x^*)=0\Leftrightarrow x^* = 1-\frac \gamma\lambda \cdot \frac {p+\lambda \, \fdp{N}\, \umin }{p+\lambda\,(\fdp{N}-1)\, \umin}
$$
%\Leftrightarrow 0=\lambda(1-x)(1-\frac{ u_{min}\lambda}{p+n\lambda u_{min}})-\gamma \\
%\Leftrightarrow \gamma=\lambda(1-x)(\frac{ p+(n-1)u_{min}\lambda}{p+n\lambda u_{min}})\\
%\Leftrightarrow (\frac{\gamma(p+n\lambda u_{min})}{\lambda( p+(n-1)u_{min}\lambda)})=(1-x)\\
%\Leftrightarrow 1-\frac{\gamma(p+n\lambda u_{min})}{\lambda( p+(n-1)u_{min}\lambda)}=x
%\end{array}
%$$

Now we can assume  for player $i$ a switch occurs in $x^*$. Hence, we impose the continuity of  the value function \cite{BaCaDo}. Because
it is continuous on both sides of the threshold $x^*$, the limit values $V_i(x^-)$ for $x\uparrow x^*$ when $(u_i,a_{-i})=(\umax,(n-1)\umax)$ 
and $V_i(x^+)$ for $x\downarrow x^*$ when $(u_i,a_{-i})=(\umin,(n-1)\umin)$ need to be the same. This will determine constant $K^*$. The equation to be solved is thus 
\begin{eqnarray}\label{esysK}
&& \hskip-7mm \small K^* (1-x^*)^{-\frac {p}{\lambda n \umax}}
%&&- \frac{p \lambda \umax(1-x) + (p+\lambda n \umax)\gamma(\umax-\umin)}{p(p+\lambda n \umax)} \no \\
- \frac{\lambda \umax (1-x^*) }{p+\lambda \fdp{N} \umax}- \frac \gamma{p}(\umax-\umin) \no \\
&&= \small - \frac{ \umin\lambda (1-x^*)}{p+\fdp{N} \lambda \umin}
\end{eqnarray}
and the expression for $K^*$  writes as in the statement. Finally, we observe from \eqref{esysK} that indeed $K>0$: in fact 
$$
\frac{ \umax\lambda}{p+\fdp{N}\lambda \umax} > \frac{ \umin\lambda}{p+\fdp{N}\lambda \umin}
$$ 
which concludes the proof.
\end{IEEEproof}

\bibliographystyle{IEEEtran}
\bibliography{kbib}  % sigproc.bib is the name of the Bibliography in this case

\end{document}